\documentclass[aps]{revtex4}
\usepackage{graphicx}
\usepackage{dcolumn}
\usepackage{bm}
\usepackage{rotate}
\usepackage{epsfig}
\newcommand \be  {\begin{equation}}
\newcommand \bea {\begin{eqnarray} \nonumber }
\newcommand \ee  {\end{equation}}
\newcommand \eea {\end{eqnarray}}
\begin{document}
\title{Critical fluctuations and breakdown of Stokes-Einstein relation in the Mode-Coupling Theory of glasses}
\author{Giulio Biroli$^1$, Jean-Philippe Bouchaud$^{2,3}$}
\email{biroli@cea.fr,jean-philippe.bouchaud@cea.fr}
\affiliation{
$^1$ Service de Physique  Th{\'e}orique Centre d'{\'E}tudes de Saclay \\ 
  Orme des
Merisiers, 91191 Gif-sur-Yvette Cedex, France\\
$^{2}$ Service de Physique de l'{\'E}tat Condens{\'e},
Orme des Merisiers,
CEA Saclay, 91191 Gif sur Yvette Cedex, France.\\
$^3$ Science \& Finance, Capital Fund Management, 6-8 Bd
Haussmann, 75009 Paris, France.
}
\date{\today}


\begin{abstract}
We argue that the critical dynamical fluctuations predicted by the mode-coupling theory ({\sc mct}) 
of glasses provide a natural mechanism to explain the breakdown of the Stokes-Einstein relation. This breakdown, 
observed numerically and experimentally in a region where {\sc mct} should hold, is one of the major difficulty of
the theory, for which we propose a natural resolution based on the recent interpretation of the {\sc mct} transition
as a {\it bona fide} critical point with a diverging length scale. We also show that the upper critical dimension
of {\sc mct} is $d_c=8$.
\end{abstract}

\maketitle

Mode Coupling Theory ({\sc mct}) provides a useful theoretical framework to account for many of the empirical properties
of liquids, at least in the weakly supercooled region \cite{Gotze,KA}. There are however a number of 
well known difficulties with the theory,
the most prominent being the absence, in real systems, of the dynamical arrest singularity predicted by {\sc mct}. This
very singularity underlies most of the quantitative predictions of {\sc mct}, such as the divergence of the relaxation time, the
power-law shape of the correlator in the $\beta$ region, or the critical behaviour of the non-ergodic (Edwards-Anderson) 
parameter $q$ \cite{Gotze}. Additional activated relaxation processes, not described by {\sc mct}, have to be invoked to argue that the
{\sc mct} transition temperature $T_c$ should be understood as a cross-over rather than a true singularity. 
In the region between $T_c$ and the so-called onset temperature, $T_{onset}$, where the
slow dynamics regime sets in \cite{Shastry},  {\sc mct} fares quite well to describe a host of experimental and 
numerical results with one noticeable exception: the breakdown of the Stokes-Einstein relation.
This breakdown is one of the most important aspect of the
phenomenology of supercooled liquid. The
Stokes-Einstein relation states that the product of the viscosity $\eta$ 
times the self-diffusion constant $D$, divided by the temperature is a
constant independent of temperature. This relation is well obeyed in liquids at high
enough temperature.
Instead, in supercooled liquids $D \eta/T$ increases quite shaply below $T_{onset}$ \cite{BADM} 
and eventually reaches values of the order of 
$1000$ or more at the glass temperature $T_g$ \cite{Ediger}.
In other words, the self-diffusion of particles becomes much faster than 
structural relaxation. The decoupling between viscosity and diffusion is interpreted by many to 
be a direct piece of evidence for dynamical heterogeneities in glassy liquids \cite{Ediger}. 
Intuitively, this decoupling comes about 
because diffusion is dominated by the fastest particles whereas structural 
relaxation is dominated by the slowest regions \cite{Ediger}. 
Detailed explanations based on different theoretical approaches have been put
forward in the literature \cite{Stillinger,Tarjus,Wolynes,GC}.  
In most models, the breakdown of the Stokes-Einstein relation is in fact a direct measure of the 
width of the distribution of local relaxation times (see below and \cite{Ediger,Stillinger,Tarjus,Wolynes,GC}).
Standard {\sc mct} does not account for the sharp increase of $D \eta/T$
between $T_{onset}$ and $T_c$ although {\it this is precisely the region where
 {\sc mct} is supposed hold}.
The aim of this short contribution is to discuss an
explicit mechanism that leads to such a breakdown within the mode-coupling theory of
glasses, once critical fluctuations are taken into account.

In its usual interpretation, {\sc mct} describes homogeneous dynamics and neglects all fluctuations; it therefore
seems that any effect associated to dynamical heterogeneities is outside of the scope of {\sc mct}. This pessimistic 
view was however recently argued to be erroneous \cite{FP,BB,BBKR,FM} (see
\cite{KW,KT} for earlier insights). {\sc mct} should in fact be seen as a standard mean-field 
theory, except that it provides a self consistent equation for the two-body dynamical correlation function 
(describing density fluctuations) 
rather than for a one body order parameter, such as the magnetisation in the usual Curie-Weiss theory of magnets. In the 
latter case, we know that the appearance of a non trivial solution of the mean field equation below a certain critical 
temperature is in fact associated with the divergence of the magnetic susceptibility, itself related to magnetisation 
fluctuations. These fluctuations are harmless in high enough dimensions, but become dominant in dimension space less 
than four, where all critical properties of the phase transition are strongly affected by these fluctuations. 
The same scenario, although more involved, also holds for {\sc mct} \cite{BB}: above a certain dimension $d_c$, {\sc mct} 
is expected to be quantitatively accurate (at least in a regime where the above mentioned activated events can be neglected), 
whereas for $d < d_c$, dynamical fluctuations play a major role and must be properly accounted for. In this case, 
fluctuations of the (two body) dynamical correlation are measured through a {\it four body} 
correlation $G_4(\vec r,t)$; its integral over space define, in analogy with ferromagnets, a dynamical susceptibility 
called $\chi_4(t)$ in the recent literature \cite{Glotzer,Berthier,TWBBB}. 
The extension of {\sc mct} to compute $G_4(\vec r,t)$ reveals that $\chi_4$ indeed diverges 
as the mode-coupling temperature $T_c$ is approached \cite{BB,BBKR}. This indicates that the dynamics becomes 
correlated over larger and
larger length scales as the system freezes, which is in fact expected on general grounds: a diverging relaxation time should 
necessarily involve an infinite number of particles \cite{MS}. The growth of $\chi_4$ has recently been detected in numerical 
simulations and experimentally in supercooled liquids 
\cite{Glotzer,Berthier,Science} and in granular media \cite{DMB}; 
it should also transpire in the divergence of the 
non-linear susceptibility of glassy systems \cite{BBchi3}. 

As discussed in \cite{BB} 
the spatial correlations underlying the mode-coupling singularity necessarily lead to the existence of 
an upper critical dimension $d_c$ for this problem. In \cite{BB} it was shown
that in the early $\beta$-relaxation regime, the four-point correlator 
reads, in Fourier space, $\hat G_4(\vec k,t \approx \tau_\beta) \sim (k^2 + \sqrt{\epsilon})^{-1}$, where $\epsilon=(T-T_c)/T_c$. 
From this, one can estimate, in the spirit of a Ginzburg argument \cite{BB}, the fluctuations of the non-ergodic parameter 
$q$ in a region of size $\xi \sim \epsilon^{-1/4}$ to find $\xi^d \delta q \sim \xi^{d+2/2}$. Imposing that $\delta q$ 
must be much smaller than the critical behaviour predicted by {\sc mct}, i.e. $q_c - q \sim \sqrt{\epsilon}$, one finds 
that this assumption is only consistent when $d > d_c=6$. For $d < 6$, the fluctuations become dominant for $\epsilon$ small 
enough and change all the critical exponents. But most interestingly, as we will argue below, these fluctuations are also responsible 
for the breakdown of the Stokes-Einstein relation. Before doing so, we should however point out that the analysis of \cite{BB} in fact 
overlooked a contribution due to the coupling between dynamic fluctuations and slow conserved degrees of freedom such as density and 
energy fluctuations \cite{BBBKMR}. These fluctuations, {\it close enough to the transition}, become the leading ones and
change the above mean-field critical behaviour of $\hat G_4$ to $\hat G_4(\vec k,t \approx \tau_\beta) \sim (k^2 +
\sqrt{\epsilon})^{-2}$ \cite{BBBKMR}. In this case, the fluctuations of the non-ergodic parameter in a region of size 
$\xi$ grow as $\xi^{d+4/2}$, which becomes dominant below the critical dimension $d_c=8$\footnote{The analysis of \cite{BB} is 
strictly speaking correct for models that are characterized by a dynamics without conserved quantities. An example is a finite 
dimensional p-spins model for which the upper critical dimension is therefore $d_c=6$.}. A diagrammatic derivation of this upper critical
dimension is presented in the Appendix. Note that in reality the situation is more complicated: from numerical simulations of 
Lennard-Jones systems \cite{BBBKMR} one finds that fluctuations due to slow conserved degrees of freedom only become dominant very close to the 
transition, where {\sc mct} breaks down. For colloids, where {\sc mct} fares rather well, we do not have yet quantitative estimates
of the relative contribution of these fluctuations. It might therefore well be that in cases of experimental or numerical interest, 
the original analysis of BB holds and the {\it effective} upper critical dimension is $d_c=6$. In any case, the aim of this paper is not 
to obtain quantitative predictions but only to showing that critical dynamical {\sc mct} fluctuations provide a natural scenario for the
decoupling of self-diffusion and viscosity.

In principle, the contribution of critical fluctuations to the viscosity and diffusion constant should be calculated 
using a renormalisation group around $d_c$. Here, we do not attempt to do this but argue physically that such a 
program should lead to a decoupling between these two quantities. We assume that non trivial critical exponents 
describe, in $d < d_c$, the different physical quantities; for example $\xi \sim \epsilon^{-\nu}$, 
$q_c - q \sim \epsilon^\beta$, $\tau \sim \epsilon^{-\gamma}$ and $G_4(\vec r) \sim g(r/\xi)/r^{d-4+\eta}$. Mode-coupling
theory describes how the random potential created by density fluctuations slows down the particles. 
Close to the {\sc mct} transition, this self-generated random potential persists on a timescale comparable to the one responsible 
for structural relaxation. Hence, it acquires a static component that self-consistently traps the particles at the
{\sc mct} transition, where the relaxation timescale diverges. 
The strength of this self-generated component, and hence of the trapping potential, is measured by the plateau
value $q$ of the correlation function, which therefore acts as an effective 
coupling constant. When $q$ reaches $q_c$, the random potential is sufficiently strong to prevent the particles from moving. 
Clearly, activated effects are expected to destroy the transition (any finite potential barrier can be overcome in finite time), 
but within {\sc mct}, these processes are neglected. 

The local difference between $q$ and $q_c$ therefore plays a major role: if the local value of $q$ is slightly smaller, particles 
are less trapped and diffusion is enhanced. If, on the other hand, the local value of $q$ is slightly larger, particles are 
frozen\footnote{This means that the local Debye-Waller factor is expected to be strongly correlated by the particles' 
propensity to diffuse, as indeed recently found numerically in \cite{HarrNew}, and L. Berthier, unpublished.}. In other words, 
the self-generated disorder fluctuates in space and leads to fluctuations of the critical temperature  
on correlated regions of linear size $\xi$\footnote{see \cite{Tcfluct} for a discussion of the role of critical temperature 
fluctuations in disordered systems.}. Since the (local) relaxation time is set by the distance from the (local)
critical temperature, the dynamics strongly fluctuates from one correlated region to another.
This is precisely why the fluctuations
of the local correlation function, measured by $G_4(\vec r,t)$, play a crucial role to generate strong dynamical 
heterogeneities. Parallel to the Ginzburg argument, the fluctuations (per
particle) of the
non-ergodic parameter in a region of size 
$\xi \sim \epsilon^{-\nu}$ are given by $\delta q \sim \xi^{(4-d-\eta)/2} \sim \epsilon^{-\nu(4-d-\eta)/2}$, to be compared 
with $q_c - q \sim \epsilon^\beta$. These fluctuations of the effective coupling constant in turn induce local 
fluctuations of the relaxation time as:
\be
\frac{\delta \tau}{\tau} = \frac{\gamma}{\beta} \frac{\delta q}{q_c-q} \sim \epsilon^{\frac{\nu(d+\eta-4)}{2}-\beta}. 
\ee
For $d > d_c$, the mean field value of the exponents $\nu=1/4$, $\beta=1/2$ can be used, and one finds that 
${\delta \tau}/{\tau} \to 0$ close to the {\sc mct} transition: dynamical heterogeneities are mild and do not jeopardize 
the Stokes-Einstein relation, as predicted by {\sc mct} calculations that neglect spatial fluctuations altogether. 
Interestingly, this conclusion does not hold for $d < d_c$, where the relative width of the local relaxation time 
distribution {\it must increase} as $\epsilon \to 0$. For $\epsilon$ not too small, one can still use the mean-field value 
of the exponents, but now the fluctuations of local relaxation times are found to {\it increase} as 
${\delta \tau}/{\tau} \sim \epsilon^{-(d_c-d)/4}$, before one enters the Ginzburg region, where fluctuations eventually change the 
values of the exponents. For standard phase transitions, the hyperscaling
relation $\nu(d+\eta-2)-2\beta=0$ is such that fluctuations remain of order one when
$\epsilon \to 0$. Its generalization to the {\sc mct} transition would be 
$\nu(d+\eta-4)-2\beta=0$. However, it is not clear whether within {\sc mct} hyperscaling holds since the {\sc mct} 
transition is purely dynamical, and has a mixed first order/second order nature ($q_c$ is non zero at $T=T_c$ while the relaxation 
time is diverging as a power law). A detailed renormalisation group study of the replica field theory with a cubic term could help 
settling this interesting technical point. In any case, we
expect that in three dimensions, neglecting the role of the activated processes (that cut off the transition)
and entering in the regime where the dynamics is supposed to slow down
because of the {\sc mct} critical point (i.e. between $T_{onset}$ and $T_c$)  ${\delta \tau}/{\tau}$ should first grow as $\epsilon$ 
decreases, and either saturate or diverge as $\epsilon \to 0$, depending on the validity of hyperscaling.  

As already discussed, a key ingredient that generates violation of the
Stokes-Einstein relation is the fluctuation of the local relation timescale $\tau$. 
Different models capture this effect. For instance, assuming as in
\cite{Tarjus} local fluctuations of the diffusion coefficient one finds that 
the Stokes-Einstein violation is related to the ratio between $\langle 1/\tau
\rangle $ and $1/\langle \tau \rangle$. Another model consists in assuming
that particles hop with random relaxation times. In this case the self
diffusion coefficient is given by $D = \ell^2/\langle \tau \rangle$, 
where $\ell$ is the typical hopping distance, expected to be of the order of the 
particle size and roughly temperature independent. The viscosity, or the terminal relaxation time, on the other
hand, are given by the integral of the average correlation function, which weighs slow regions proportionally to the 
local relaxation time; therefore $\eta \sim \langle \tau^2 \rangle/\langle
\tau \rangle$. In both models the crucial ingredient to evaluate violation of
the Stokes-Einstein relation are indeed local fluctuations of the relaxation
times. For example for the latter model, one gets:
\be\label{eq1}
\frac{D \eta}{K_{SE}T} \propto 1 + \langle (\delta \tau)^2 \rangle/\langle \tau \rangle^2 \sim 
1 + C \epsilon^{\nu(d+\eta-4)-2\beta},
\ee
where $C$ is a numerical constant of order unity and $K_{SE}$ is the high
temperature value of $D \eta/T$. For $d < d_c$, this formula predicts an
increasing decoupling between viscosity and diffusion between $T_{onset}$ and $T_c$. 
If $C$ is not too small, we predict a power-law viscosity-diffusion 
decoupling as $T_c$ is approached, with exponents that change as one enters the Ginzburg region. Such a  
behaviour may have been observed in numerical simulations of Lennard-Jones particles \cite{KA}, where $D \eta/T$
is found to increase by a factor $\sim 5$ in the {\sc mct} region \cite{BADM}, while {\sc mct} predicts an increase
of a mere $1.2$ \cite{KA,Voigt}. These results are interesting because they show that 
both the viscosity and the inverse diffusion constant can be made to 
scale relatively well with $\epsilon=(T-T_c)/T_c$ with the same $T_c$ for both quantities, but with different exponents. 
In other words, this scenario is compatible with a `fractional' Stokes-Einstein relation, $D \sim \eta^{-k}$ with $k < 1$. More 
precisely, the numerical results are $\eta \sim \epsilon^{-2.4}$, whereas $D \sim \epsilon^{1.8}$, so that 
$D \eta \sim  \epsilon^{-0.6}$ (see \cite{BADM} and refs therein). These results suggest that the observed breakdown of the 
Stokes-Einstein relation might  
indeed be related to critical fluctuations. A similar breakdown of the Stokes-Einstein relation {\it in the mode coupling
region} has also been observed in hard sphere systems, both experimentally \cite{Bonn} and numerically \cite{Voigt,Sz}. 
Of course, much stronger violations are observed closer to $T_g$, but outside the region where {\sc mct} can be valid. These 
should be related to the broadness of {\it activated} relaxation time distributions and are outside the scope of 
the present discussion \footnote{Note however that activated events play a role at long times even above $T_c$, and 
might thus be at least partly responsible for the decoupling \cite{Denny,Heuer}.}.

Summarizing this short contribution, we have argued that a dramatic consequence, in low space dimensions, of the critical dynamical 
fluctuations 
predicted by the mode-coupling theory of glasses is the breakdown of the Stokes-Einstein
relation. This breakdown, observed numerically and experimentally in a region where {\sc mct} should hold, i.e. above 
$T_c$, is one 
of the major difficulty of {\sc mct} (setting aside other well known difficulties related to activated processes), 
and for which we provide a natural interpretation.
We believe that this matter deserves more investigation, both 
\begin{itemize}
\item theoretically, to estimate the value of the exponents 
describing this breakdown in $d < d_c$ and analyze the interplay between 
activated processes and critical dynamical fluctuations; 
\item and numerically and/or experimentally to ascertain the connection between dynamical 
fluctuations and viscosity-diffusion decoupling in the $T > T_c$ region, for example by comparing the location 
$\tau$ and the width $\delta \tau$ of the peak of the four-point susceptibility $\chi_4(t)$. 
\end{itemize}
Numerically one could also test the 
dimension dependence of Stokes-Einstein violation, and compare with other scenarii, such as suggested by Kinetically
Constrained Models which also predict such violations \cite{KCM}. If confirmed, those effects would be a direct 
proof of the importance of critical fluctuations in glasses. This would put the program of extending {\sc mct} using 
renormalisation group methods on top of the agenda. 

\vskip 1cm

We thank L. Berthier, D. R. Reichman, R. Richert and M. Wyart for many useful discussions on these topics.
GB is partially supported by EU contract HPRN-CT-2202-00307 (DYGLAGEMEM).

\section*{Appendix}
In the following we shall show that below eight dimensions there are infrared divergences that change 
the mean-field scaling derived in BB and \cite{BBBKMR}.

The natural framework to analyze critical dynamical fluctuations and their role
in determining the upper critical dimension is dynamical field theory.
The starting point to analyze the dynamics of super-cooled liquids consists in
writing down some exact or phenomenological stochastic equations for the evolution of the slow conserved
degrees of freedom. Field-theories are obtained through the
Martin-Siggia-Rose-deDominicis-Janssen method, where one first introduces
response fields enforcing the correct time evolution and then 
averages over the stochastic noise~\cite{Zinn-Justin}. 
Standard {\sc mct} corresponds to a self-consistent one-loop approximation (see Refs.~\cite{ABL,das,MR} for a 
discussion of the different field-theories and subtleties appearing in
the derivation of {\sc mct}).

\begin{figure} 
\psfig{file=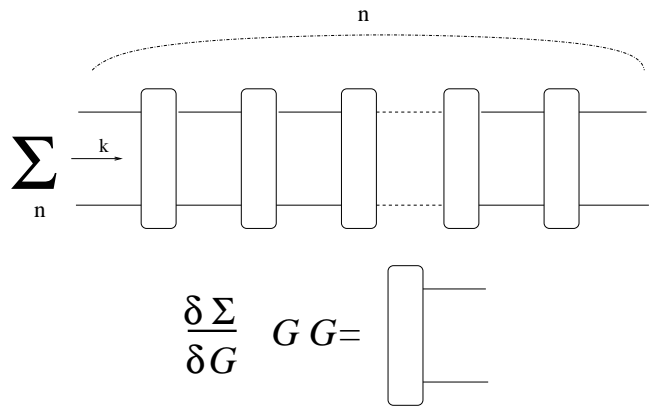,width=8.5cm} 
\caption{\label{ladders} 
Diagrammatic representation of the parquet diagrams. $\vec k$ denotes 
the wave-vector entering into the ladders.} 
\end{figure}

\begin{figure} 
\psfig{file=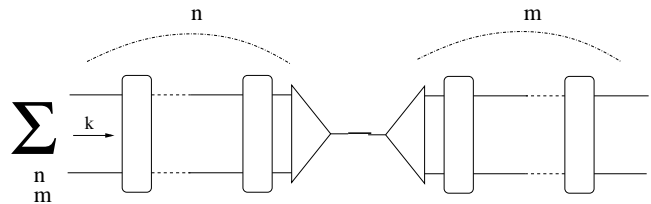,width=8.5cm} 
\caption{\label{laddersquared} 
Diagrammatic representation of the squared-parquet diagrams. $\vec k$ denotes 
the wave-vector entering into the ladders. 
} 
\end{figure}

Here we do not provide any detail and we refer to \cite{ABL,BBBKMR} for a thorough presentation.
We only recall that generically the four-point function can be written  
in terms of the so called parquet diagrams, see \cite{BlaizotRipka} for a general
introduction and \cite{BBBKMR} for an application to supercooled liquids.
As shown in \cite{BBBKMR} the four-point function can be exactly expressed as a sum of
parquet and ``squared-parquet'' diagrams (we will use the same notation of \cite{BBBKMR}), see
Fig. \ref{ladders} and Fig. \ref{laddersquared}.

The elementary block used to construct the parquet diagrams corresponds to $[\delta \Sigma/\delta G] GG$ where 
$\Sigma$ is the self-energy considered as a function of the dressed propagator $G$, see \cite{BBBKMR}. 
{\sc mct} consists in retaining only the bubble diagram for the self-energy. In this case the
elementary block is very
simple and leads to the ladder diagrams studied in BB and \cite{BBBKMR}, which lead to the mean-field 
scaling discussed in the main text. The diagram in Fig. \ref{ladders} gives a contribution to $G_4(k,t)$ that, for times $t$ 
in the $\beta$ regime and for small $k$, has a critical behavior like $1/(k^2+\sqrt{\epsilon})$. The diagram in 
Fig. \ref{laddersquared}, on the other hand, gives a contribution that has a critical behavior like $1/(k^2+\sqrt{\epsilon})^2$.

From the point of view of critical phenomena, one expects that adding diagrams other than the bubbles
to the self-energy should be harmless and should not change the scaling for all dimensions larger 
than the upper critical dimension $d_c$. In order to show that this is the case and determine 
$d_c$ one has to prove that (1) adding any type of diagram to the bubbles does not change the {\sc mct}
(mean-field) critical properties in presence of an infrared cut-off on momenta integration and (2) 
equating the infrared cut-off to zero leads to a change of the mean-field scaling only 
below the upper critical dimension $d_c$. The first part of this procedure is lengthly and difficult 
and it will be shown elsewhere \cite{ABB}. Concerning the second part, we shall only show an example of 
diagrams that are responsible for changing the mean-field scaling below eight dimension. (A full 
analysis showing that above eight dimensions all diagrams are harmless is outside the 
scope of this article and is left for future work).

\begin{figure} 
\psfig{file=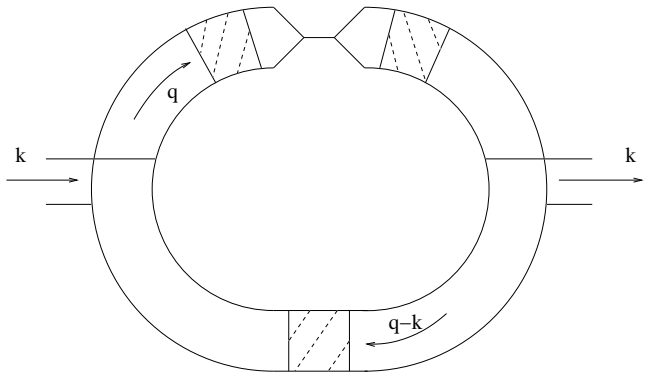,width=8.5cm} 
\caption{\label{divergentdiagram} 
Divergent diagram below eight dimension assuming mean-field scaling.
One box represents the parquet diagram shown in Fig. \ref{ladders}.
Note that the lines on the left have to be considered without propagators.  
} 
\end{figure}

In order to show an example of divergent diagram below eight dimension one has to focus on the 
elementary block $[\delta \Sigma/\delta G] GG$ used to construct the parquets. The small $k$ behavior
of this block determines the {\sc mct} mean-field scaling discussed above ($k$ is the momentum entering into the block as shown in 
Fig. 1 and 2). Within {\sc mct}, and more generally,
if one puts an infrared cutoff, one finds a scaling $k^2+\sqrt{\epsilon}$ at small $k$. One can show that
this assumption is inconsistent in low enough dimension, focusing on an explicit diagram, see Fig. 
\ref{divergentdiagram} \footnote{Note that the diagram shown in Fig. \ref{divergentdiagram} is composed of a 
double parquet on the top and 
a single parquet on the bottom. It is not possible to have double parquet both on the top and bottom parts because 
joining the two lines on the right and on the left one must construct a two particle irreducible diagram,
and this diagram could be cut in two pieces just cutting the two middle lines. }.

In order to do that, we shall assume that the mean-field scaling is correct and  
compute the small $k$ behavior of this diagram. In dimensions less than $d_c$, the corresponding contribution to 
$[\delta \Sigma/\delta G] GG$ at small $k$ will be found to differ from mean-field. 
For the diagram in Fig. \ref{divergentdiagram}, the small $k$ dependence is given by 
(we omit the dependence on the non-critical wave-vectors):
\[
I(k,\epsilon)\sim \int \frac{1}{(q^2+\sqrt{\epsilon})^2} \frac{1}{(q-k)^2+\sqrt{\epsilon}}d^dq
\]  
For $\epsilon=0$ the small $k$ behavior is then given by $I(0,0)+ c k^2+...$ where $I(0,0)$ and $c$
are two well-defined constants above eight dimensions (that can be explicitely computed from $I(k,\epsilon)$).
$I(0,0)$ and $c$ lead, respectively, to a renormalization of the critical temperature and of the ``rigidity'' constant
in front of the $k^2$ term in $G_4$. However, below eight dimensions, $c$ diverges (if the infrared cutoff vanishes) 
and the behavior of $I(k,0)$ is singular at small $k$: $I(0,0)+c' k^{2-(8-d)}+...$. 
This clearly shows that the assumption of mean field scaling is not correct. In order to go further
and compute the non mean field exponents one should perform a renormalization group treatment, or some 
self-consistent (mode-coupling!) resummation of these dangerous diagrams. We leave this for further investigations.

As a conclusion we find, in agreement with the simple Ginzburg argument developed in the main text,
that below eight dimension critical fluctuations are not governed by mean-field theory. 

Finally, we remark that for a system with no conserved degrees of freedom, as it is the case for the
Langevin dynamics of disordered p-spin systems, the squared-parquet diagrams are absent.
Therefore the divergent diagram corresponding to the one in Fig. \ref{divergentdiagram} is now formed by
single parquets both on the top and bottom parts of the diagram. Repeating the same analysis performed above, 
one finds $d_c=6$ in this case, as found in BB.


\begin{thebibliography}{99}
\bibitem{Gotze} W. G{\"o}tze, L. Sj{\"o}gren, Rep. Prog. Phys. {\bf 55} 241 (1992); 
W. G{\"o}tze, Condensed Matter Physics, {\bf 1}, 873 (1998); W. G{\"o}tze,
J. Phys.: Condens. Matter {\bf 11} A1-A45 (1999).

\bibitem{KA} W. Kob, H. C. Andersen, Phys. Rev. E {\bf 51} 4626 (1995); {\bf 52} 4134 (1995); see also: W. Kob,
in {\it Slow relaxations and non-equilibrium dynamics in condensed matter}, Les Houches,
Session LXXVII, J. L. Barrat, M. Feigelman, J. Kurchan, J. Dalibard Edts, Springer-EDP Sciences (2003).
\bibitem{Shastry} S. Sastry, P. G. Debenedetti and F. H. Stillinger, Nature
  {\bf 393}, 554 (1998).
\bibitem{BADM} P. Bordat, F. Affouard, M. Descamps and F. Muller-Plathe,
  J. Phys. Cond. Matt. {\bf 15} 5397 (2003).
\bibitem{Ediger} see e.g. M. D. Ediger, Ann. Rev. Phys. Chem. {\bf 51}, 99 (2000).
\bibitem{Stillinger} F. H. Stillinger, J. Chem. Physics {\bf 89}, 6461 (1988); 
F. H. Stillinger and J. A. Hodgdon Phys. Rev. E {\bf 50} 2064 (1994).
\bibitem{Tarjus} G. Tarjus, D. Kivelson, J. Chem. Phys. {\bf 103} 3071 (1995). 
\bibitem{Wolynes} X. Xia and P. G. Wolynes, J. Phys. Chem. B, {\bf 105}, 6570 (2001).
\bibitem{GC} Y. Jung, J.P. Garrahan and D. Chandler, Phys. Rev. E {\bf 69}, 061205 (2004). 
\bibitem{FP} S. Franz, G. Parisi, J. Phys.: Condens. Matter {\bf 12}, 6335 (2000); C. Donati, S. Franz, G. Parisi, 
S. C. Glotzer, J. Non-Cryst. Sol., {\bf 307}, 215-224 (2002).
\bibitem{BB} G. Biroli, J. P. Bouchaud, {\it Europhys. Lett.} {\bf 67}, 21-27 (2004). 
\bibitem{BBKR} G. Biroli, J.P. Bouchaud, K. Miyazaki, D. R. Reichman, {\it Inhomogeneous Mode-Coupling Theory and 
growing dynamic length in supercooled liquids}, cond-mat/0605733.
\bibitem{FM}  S. Franz, A. Montanari, {\it Dynamical and mosaic length scales in a Kac glass model}, cond-mat/0606113.
\bibitem{KW} T.R. Kirkpatrick, P.G. Wolynes, Phys. Rev. B {\bf 36}, 8552 (1987).  
\bibitem{KT} T.R. Kirkpatrick and D. Thirumalai, Phys. Rev. A {\bf 37}, 
4439 (1988).  
\bibitem{Glotzer} C. Bennemann, C. Donati, J. Bashnagel, S. C. Glotzer, Nature, 
{\bf 399}, 246 (1999), S. C. Glotzer, J. Non-Cryst. Solids {\bf 274}, 342 (2000).
\bibitem{Berthier} L. Berthier, {\it Phys. Rev. E} {\bf  69}, 020201/1-4 (2004); 
S. Whitelam, L. Berthier, J.P. Garrahan, {\it Phys. Rev. Lett.} {\bf 92}, 185705/1-4 (2004).
\bibitem{TWBBB} C. Toninelli, M. Wyart, G. Biroli, L. Berthier, J.P. Bouchaud, 
{\it Phys. Rev. E} {\bf 71}, 041505/1-20 (2005).
\bibitem{MS} A rigorous proof for a large class of glassy systems has been
    presented in A. Montanari, G. Semerjian, {\it  Rigorous Inequalities
    between Length and Time Scales in Glassy Systems}, cond-mat/0603018.  
\bibitem{Science} L. Berthier, G. Biroli, J.-P. Bouchaud, L. Cipelletti, D. El Masri, D. L'Hote, 
F. Ladieu, M. Pierno, Science {\bf 310}, 1797 (2005).
\bibitem{DMB} O. Dauchot, G. Marty and G. Biroli, Phys. Rev. Lett. {\bf 95},
265701 (2005).  
\bibitem{BBchi3}  J. P. Bouchaud, G. Biroli, Phys. Rev. B {\bf 72} 064204 (2005).
\bibitem{BBBKMR}  L. Berthier, G. Biroli, J.P. Bouchaud, W. Kob, K. Miyazaki,
  D. R. Reichman, cond-mat/0609656 and 0609658.
\bibitem{HarrNew} A. Widmer-Cooper and P. Harrowell, {\it Predicting the long time dynamic heterogeneity in a supercooled 
liquid on the basis of short time heterogeneities}, e-print cond-mat/0512035.
\bibitem{Tcfluct} A. Aharony and A.B. Harris, Phys. Rev. Lett. {\bf 77} 3700
  (1996); A. Aharony, A.B. Harris and S. Wiseman Phys. Rev. Lett. {\bf 81} 252
  (1998); S. Wiseman and E. Domany, Phys. Rev. Lett. {\bf 81} 22 (1998).
\bibitem{Bonn} D. Bonn, W. Kegel, J. Chem. Phys. {\bf 118} 2005 (2003)
\bibitem{Denny} R. A. Denny, D. R. Reichman, and J. P. Bouchaud, Phys. Rev. Lett. {\bf 90} 025503 (2003).
\bibitem{Heuer} B. Doliwa, A. Heuer, Phys. Rev. E {\bf 67} 030501(R) (2003)
\bibitem{Voigt} Th. Voigtmann, A. M. Puertas, M. Fuchs, Phys. Rev. E {\bf 70}, 061506 (2004).
\bibitem{Sz} S. Kumar, G. Szamel, J. Douglas, {\it Nature of the Breakdown in
    the Stokes-Einstein Relationship in a Hard         
Sphere Fluid}, cond-mat/0508172.
\bibitem{KCM} see e.g. Y. Jung, J.P. Garrahan and D. Chandler, Phys. Rev. {\bf 69} 061205 (2004); preprint
cond-mat/0504535; L. Berthier, D. Chandler, and J.P. Garrahan, Europhys. Lett. {\bf 69} 230 (2005).
\bibitem{ABB}A. Andreanov, G. Biroli, J.-P. Bouchaud, {\it in preparation}.
\bibitem{Zinn-Justin} J. Zinn Justin, {\it Quantum field 
theory and critical phenomena} (Oxford University Press, Oxford, 2002).
\bibitem{ABL} A. Andreanov, G. Biroli, and A. Lef{\`e}vre,
J. Stat. Mech. P07008 (2006).
\bibitem{das} 
S.~P. Das, Rev. Mod. Phys. {\bf 76}, 785 (2004).
\bibitem{MR}
 K. Miyazaki and D.~R. Reichman, 
 J. Phys. A: Math. Gen. {\bf 38} (2005) L343.
\bibitem{BlaizotRipka}
J.-P. Blaizot, G. Ripka,
{\it Quantum Theory of Finite Systems}, Editions Phenix, Kiev, 1998.

\end{thebibliography}
\end{document}